\documentclass[aps,twocolumn,showpacs]{revtex4}
\usepackage{graphicx}% Include figure files
\usepackage{bm}% bold math
\usepackage{amsmath}
\begin{document}

\title{Vortex Charges in High Temperature Superconductors}

\author{Yan Chen$^1$ Z. D. Wang$^{1,2}$, Jian-Xin Zhu$^3$ and C. S. Ting$^1$}
\affiliation{$^1$Texas Center for Superconductivity and Department
of Physics, University of Houston, Houston, TX 77204\\
$^2$Department of Physics, University of Hong Kong, Pokfulam Road,
Hong Kong, China\\ $^3$Theoretical Division, MS B262, Los Alamos
National Laboratory, Los Alamos, NM 87545}

\begin{abstract}
Based on an effective model Hamiltonian with competing
antiferromagnetic (AF) and $d$-wave superconductivity (DSC)
interactions, the vortex charge in high $T_c$ superconductors is
investigated by solving self-consistently the Bogoliubov-de Gennes
equations.  We found that the vortex charge is always negative
when a sufficient strength of AF order is induced inside the
vortex core, otherwise, the vortex charge is positive.
% Both the hole-rich
%normal and electron-rich AF vortices are found, depending on the
%strength of induced AF order. The vortex charge is strongly influenced by the
%competing effects from the AF order and the DSC order at the
%vortex core.
By tuning the on-site Coulomb repulsion $U$ or the doping
parameter $\delta$, a transition between the positive and negative
vortex charges may occur. The vortex charge at optimal doping has
also been studied as a function of magnetic field. Recent NMR and
Hall effect experiments may be understood in terms of the present
results. New imaging experiments should be able to probe the
vortex charge directly.
\end{abstract}

\pacs{74.60.Ec, 74.20.-z, 74.25Jb}

\maketitle

The vortex structure in high temperature superconductors (HTS) has
attracted significant interest for many years. Since the parent
compounds are antiferromagnetic (AF) Mott insulators, novel
physical properties of HTS including those in the vortex state
would be expected due to the competition between spin magnetism
and superconductivity in these systems. It has been shown
theoretically~\cite{Arovas97,Demler,Zhang01,Ogata99,Machida,Zhu01,chen02,Franz}
that the AF order may appear and coexist with the underlying
vortices. In a neutron scattering experiment by Lake {\em et
al.}~\cite{Lake01}, a remarkable AF-like spin density wave (SDW)
was observed in the optimally doped La$_{2-x}$Sr$_x$CuO$_4$ in the
presence of a strong magnetic field. A moun spin rotation
measurement by Miller {\em et al.}~\cite{Miller} studied the
internal magnetic field distribution in the vortex cores of
underdoped YBa$_2$Cu$_3$O$_{6+ x}$, and it was revealed a feature
in the high-field tail which fits well to a model with static
alternating magnetic field. A very recent nuclear magnetic
resonance (NMR) experiment by Mitrovic {\em et al.}~\cite{NMR}
showed that the presence of AF order is markedly enhanced in the
vortex cores of near-optimally doped YBa$_2$Cu$_3$O$_{7- \delta}$.
These experiments have provided a strong support for the existence
of AF order inside the vortex core in appropriately doped HTS.

On the other hand, the vortex charge in superconductors has also
been paid considerable attention both
theoretically~\cite{KF95,Blatter95,Machida98,KLB01,Matsumoto} and
experimentally~\cite{Matsuda98,Yeh99,Matsuda01}. In the framework
of the BCS theory, Blatter {\em et al.}~\cite{Blatter95} pointed
out that for $s$-wave superconductor the vortex charge is
proportional to the slope of the density of states at the Fermi
level. Hayashi {\em et al.}~\cite{Machida98} proposed that the
vortex charge is always hole-like and is determined by the
quasiparticle structure which is independent of the slope of the
density of states. However, the NMR and nuclear quadrupole
resonance (NQR) measurements on YBCO ~\cite{Matsuda01} seemed to
obtain results for the vortex charge, contradictory to that
predicted from the existing BCS theory regarding to both sign and
order of magnitude. In view of this significant deviation,
together with the fact that the strong electron correlation with
the $d$-wave superconducting (DSC) pairing has not been considered
in the existing theories for the vortex charge,  we believe that
the  vortex charge in HTS should be strongly influenced by the
competition effect from the AF and DSC orders, of which the former
will play a crucial role in determining the charge nature. Also
interestingly, Hall effect experiments~\cite{Matsuda98} for HTS
seemed to indicate that the Hall signal is electron-like in the
underdoped up to slightly overdoped regime but hole-like in the
overdoped regime, which could be related to the sign of vortex
charge~\cite{KF95}. Therefore, it is important to develop a sound
theory for the vortex charge with the strong electron correlation
and the $d$-wave feature of HTS  being taken into account.

In this Letter, we shall answer two crucial questions in detail:
what is mainly responsible for the vortex charge in the HTS? and
how is the sign of vortex charge affected by the doping and the
on-site Coulomb repulsion $U$? Based on a widely adopted effective
model Hamiltonian with competing SDW and DSC orders and using a
well-developed numerical method~\cite{Wang95}, we study the vortex
charge in the mixed state of $d$-wave HTS subjected to a strong
magnetic field.  It is found that the vortex charge is mainly
determined by the competition of the AF order and the DSC order at
the vortex core, and the electronic structure of vortex core can
contain either AF order or normal state, corresponding to negative
(electron-like) or positive (hole-like) charge. By tuning $U$ or
the doping parameter, the transition between these two kinds of
vortices occurs.

Let us begin with an effective model Hamiltonian in a
two-dimensional (2D)lattice, in which both the DSC and SDW orders
are taken into account:
\begin{eqnarray}
H&=&-\sum_{i,j,\sigma} t_{ij} c_{i\sigma}^{\dagger}c_{j\sigma}
+\sum_{i,\sigma}( U n_{i {\bar {\sigma}}} -\mu)
c_{i\sigma}^{\dagger} c_{i\sigma} \nonumber \\ &&+\sum_{i,j} (
{\Delta_{ij}} c_{i\uparrow}^{\dagger} c_{j\downarrow}^{\dagger} +
h.c.)\;,
\end{eqnarray}
where $c_{i\sigma}^{\dagger}$ is the electron creation operator,
$\mu$ is the chemical potential, and the summation   is over the
nearest neighboring sites. In the presence of magnetic field $B$
perpendicular to the plane, the hopping integral can be expressed
as  $ t_{ij}= t_0 \exp[{i \frac{ \pi}{\Phi_{0}} \int_{{\bf
r}_{j}}^{{\bf r}_{i}} {\bf A}({\bf r})\cdot d{\bf r}}]$ for the
nearest neighboring sites $(i,j)$, with $\Phi_0=h/2e$ as the
superconducting flux quantum. In the presence of a strong magnetic
field, we assume the  applied magnetic field to be uniform and
choose a Landau gauge ${\bf A}=(-By,0,0)$. Since the internal
magnetic field induced by the supercurrent around the vortex core
is so small comparing with the external magnetic field that the
above assumption is justified. The two possible  SDW and DSC
orders in cuprates are defined as $\Delta^{SDW}_{i} = U \langle
c_{i \uparrow}^{\dagger} c_{i \uparrow} -c_{i
\downarrow}^{\dagger}c_{i \downarrow} \rangle$ and
$\Delta_{ij}=V_{DSC} \langle c_{i\uparrow}c_{j\downarrow}-c_{i
\downarrow} c_{j\uparrow} \rangle /2$, where $U$ and $V_{DSC}$
represent respectively the interaction strengths for two orders.
The mean-field Hamiltonian (1) can be diagonalized by solving the
resulting Bogoliubov-de Gennes equations self-consistently
\begin{equation}
\sum_{j} \left(\begin{array}{cc} {\cal H}_{ij,\sigma} &
\Delta_{ij} \\ \Delta_{ij}^{*} & -{\cal H}_{ij,\bar{\sigma}}^{*}
\end{array}
\right) \left(\begin{array}{c} u_{j,\sigma}^{n} \\
v_{j,\bar{\sigma}}^{n}
\end{array}
\right) =E_{n} \left(
\begin{array}{c}
u_{i,\sigma}^{n} \\ v_{i,\bar{\sigma}}^{n}
\end{array}
\right)\;,
\end{equation}
where the single particle Hamiltonian ${\cal H}_{ ij,\sigma}=
-t_{ij} +(U n_{i \bar{\sigma}} -\mu)\delta_{ij}$, and $n_{i
\uparrow} = \sum_{n} |u_{i\uparrow}^{n}|^2 f(E_{n})$, $ n_{i
\downarrow} = \sum_{n} |v_{i\downarrow}^{n}|^2 ( 1- f(E_{n}))$, $
\Delta_{ij} = \frac{V_{DSC}} {4} \sum_{n} (u_{i\uparrow}^{n}
v_{j\downarrow}^{n*} +v_{i\downarrow}^{*} u_{j\uparrow}^{n}) \tanh
\left( \frac{E_{n}} {2k_{B}T} \right)$, with $f(E)$ as the Fermi
distribution function and the electron density $n_{i}= n_{i
\uparrow} + n_{i \downarrow}$. The DSC order parameter at the
$i$th site is $\Delta^{D}_{i}= (\Delta^{D}_{i+e_x,i} +
\Delta^{D}_{ i-e_x,i} - \Delta^{D}_{ i,i+e_y} -\Delta^{D}_{
i,i-e_y})/4$ where $ \Delta^{D}_{ij} = \Delta_{ij} \exp[ i {
\frac{\pi}{\Phi_{0}} \int_{{\bf r}_{i}}^{({\bf r}_{i}+{\bf
r}_{j})/2 } {\bf A}({\bf r}) \cdot d{\bf r}}]$ and ${\bf e}_{x,y}$
denotes the unit vector along $(x,y)$ direction. The main
procedure of self-consistent calculation is summarized as follows.
For a given initial set of parameters $n_{i \sigma}$ and
$\Delta_{ij}$, the Hamiltonian is numerically diagonalized and the
electron wave functions obtained are used to calculate the new
parameters for the next iteration step. The calculation is
repeated until the relative difference of order parameter between
two consecutive iteration steps is less than $10^{-4}$. The
solutions corresponding to various doping concentrations can be
obtained by varying the chemical  potential.

In our calculation, the length and energy are measured in units of
the lattice constant $a$ and the hopping integral $t_0$
respectively. Magnetic unit cells are introduced where each unit
cell accommodates two superconducting flux quanta. The related
parameters are chosen as
: the DSC coupling strength is $V_{DSC}=1.2$, the linear dimension
of the unit cell of the vortex lattice is  $N_x \times N_y = 40
\times 20$. This choice corresponds the magnetic field $B \simeq
37 T$. The calculation is performed in very low temperature
regime.

Our numerical results indeed show that the AF order is absent
inside the vortex core for small $U$ and is induced when $U$
becomes larger. In Fig. 1,
\begin{figure}[b]
\includegraphics[width=8.6cm]{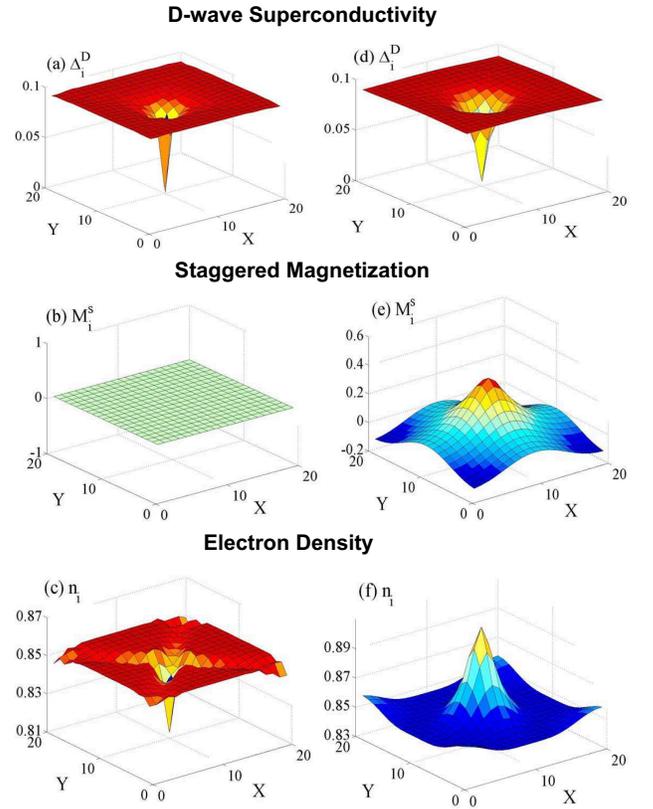}
\caption{\label{Fig1} Spatial variations of the DSC order
parameter $\Delta_{i}^{D}$ [(a) and (d)], staggered magnetization
$M_{i}^{s}$ [(b) and (e)], and net electron density $n_{i}$ [(c)
and (f)] in a $20 \times 20$ lattice. The left panels [(a), (b),
and (c)] and the right panels [(d), (e), and (f)] are for $U=2.0$
and $U=2.4$, respectively. The averaged electron density is fixed
at $\bar {n}=0.85$.}
\end{figure}
we plot typically the spatial profiles of the vortex structure for
two types of vortices: a normal $d$-wave vortex core for small $U
(=2.0)$, where the AF order is absent, and an AF core for larger
$U (=2.4)$, where the AF order is nucleated and spreads out from
the core center. They are obtained at the optimal doping
$\delta=0.15$. Panels (a)-(c) correspond to the normal core while
(d)-(f) for the AF core. Panels (a) and (d) in Fig. 1 illustrate
the DSC order parameter pattern, which vanishes at the vortex core
center.
%and recovers
%its bulk value at several coherence lengths away from the center.
The center of the vortex core is situated at site $(10,10)$.
Panels (b) and (e) display the spatial distribution of the
staggered magnetization of the induced AF-like SDW order defined
as $M_{i}^{s} =(-1)^{i} \Delta^{SDW}_{i}/U$ . No AF order is seen
in the normal core (for $U=2.0$) while the AF order exists both
inside and outside the core (for $U=2.4$) and behaves like a two
dimensional SDW with the same wavelength in the $x$ and $y$
directions. The size of the AF core here is slightly enlarged than
that of the normal core. The induced SDW order reaches its maximum
value at the vortex core center and its spatial profile retains
the same fourfold symmetry as that of the pure DSC case. The
orders of DSC and SDW coexist throughout the whole sample. The
appearance of the SDW order around the vortex cores strongly
enhances the net electron density (or depletion of the hole
density) at the vortex core as shown in panel (f). An intuitive
physical understanding of positive charge for the normal vortex
core can be given as follows: for a particle-hole asymmetric
system like doped cuprates, the chemical potential for electrons
in DSC state would be slightly lower than that of the normal
state, when a normal core is imbedded into the DSC background and
in order to reach equilibrium, electrons have to flow from the
inside to the outside of the core which leads to the electron
depletion inside the vortex core, as shown in panel (c); while in
the case of AF core, the hole number is suppressed and as a
result, the vortex carries negative charge. The enhancement of
electron number inside the AF vortex core has also been
numerically obtained by other
calculations~\cite{Machida,Zhu01,chen02,Franz}.

To examine the vortex charge $Q_v$ as functions of both $\delta$
and  $U$, the upper right inset in Fig. 2 plots the phase diagram
of $\delta$ versus $U$ for positively (hole-rich) and negatively
(electron-rich) charged vortices. It is obvious that the AF vortex
core can easily show up in the underdoped regime or with stronger
AF interaction while normal core tends to exist in the overdoped
regime or with weaker AF interaction. The electron density inside
the core is higher than the average density in the underdoped
region while the electron density becomes lower than the average
in the overdoped region. There exists a clear boundary between
these two phases. The AF order is generated in the region where
the DSC order parameter is suppressed. To estimate the core charge
of a single vortex, we first determine the vortex size by
examining the spatial profile of DSC order parameter. Next we make
a summation of the net electron density inside the vortex core. As
shown in Fig. 2, the $\delta$-dependence of $Q_v/e$ (the electron
number) for $U=2.4$ exhibits a first-order like transition at
$\delta = \delta_c$ ($ \sim 0.18$). The magnitude of the
discontinuity reduces to one third when $U=2.2$.
\begin{figure}[t]
\includegraphics[width=8.6cm]{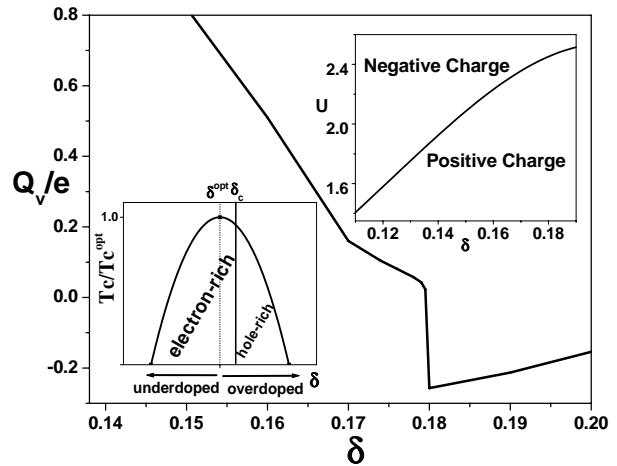}
\vspace{-0.3cm} \caption{\label{Fig2} Doping dependence of the
number of vortex charge $Q_v/e$ for $U=2.4$ where the electron
charge $e<0$. The left inset shows the doping dependence of the
sign of the vortex charge ( positive for hole-rich and negative
for electron-rich). $\delta^{opt}$ and $\delta_c$ denote
respectively the optimal doping and critical doping. The right
inset represents the phase diagram of doping level versus
interaction strength $U$ for positive and negative charged
vortex.}
\end{figure}
The critical value of the doping level $\delta_c$ is $U$-value
dependent or the sample-dependent. The larger $U$ case corresponds
to larger $\delta_c$. Recent NMR experiments~\cite{NMR} indicated
that the AF order exists in the vortex core at the optimal doping
level in cuprates,  which  may imply that the critical doping
level $\delta_c$ could be extended to slightly overdoped region.
Therefore, It is clear that the related phenomena in  the slightly
overdoped sample may be qualitatively the same as those in the
underdoped sample, e.g., the slightly overdoped sample has the
electron-rich vortex core as well. This result agrees well with
the experiment for slightly overdoped
YBa$_2$Cu$_3$O$_7$~\cite{Matsuda01}, in sharp contrast to the
hole-rich vortex core predicted by the BCS theory. Also
interestingly,  even though the origin of Hall sign anomaly is
still debatable~\cite{Wang94}, the vortex charge could make an
additional contribution to the sign change in the mixed state Hall
conductivity~\cite{KF95}. Our calculations which is schematically
shown in the lower left inset of Fig. 2 would favor that the Hall
signal is electron-like from the underdoped to slightly overdoped
regime but hole-like in the appreciable overdoped regime. This
result is consistent with the phase diagram obtained by the Hall
effect measurements~\cite{Matsuda98}.

In addition, the charge magnitude estimated from the BCS
theory~\cite{Blatter95} is two orders smaller than that of
experimental observation for HTS. The magnitude of vortex charge
estimated from our calculation is about 0.06$e$ at 22 Tesla, which
seems much larger than the experimental estimation 0.005$e$ -
0.02$e$ at 9.4 Tesla for YBa$_2$Cu$_3$O$_7$~\cite{Matsuda01}. The
reason appears to be mainly due to a much higher magnetic field
used in our calculation, which will lead to a larger AF order. The
inset of Fig. 3
\begin{figure}[t]
\includegraphics[width=8.5cm]{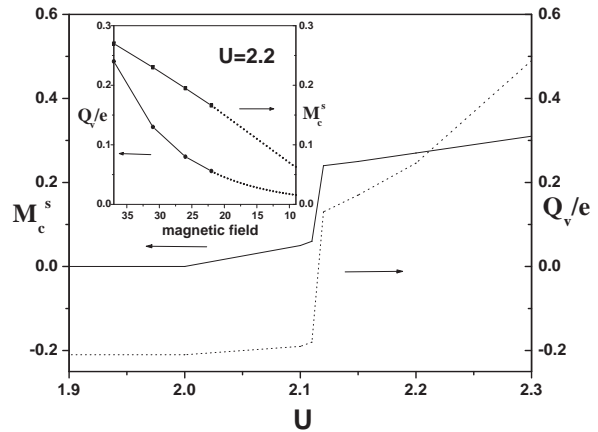}
\vspace{-0.3cm} \caption{\label{Fig3} The interaction strength
dependence of the staggered magnetization $M_{c}^{s}$ (solid line)
at the vortex center and the number of vortex charge $Q_v/e$
(dotted line). The averaged electron density is fixed at $\bar
{n}=0.85$. The inset shows the extrapolation of $Q_v/e$ and
$M_{c}^{s}$ with magnetic field strength in unit of Tesla.}
\end{figure}
represents the approximate extrapolation of the vortex charge
magnitude versus the magnetic field. The estimated vortex charge
at 9.4 Tesla is indeed in the same order of magnitude as reported
in the experiment. From Fig. 3, one can clearly see an abrupt jump
for the number of vortex charge $Q_v/e$ and staggered
magnetization at the vortex core center $M_c^s$ as $U$ varies
around 2.11, and this positively-negatively charged vortex
transition appears also to be first-order like. It is now quite
clear that the vortex charge is strongly influenced by two
competing effects --- the suppression of the DSC order at the core
center which  leads to the depletion of the electrons and the
induction of the AF order which favors the accumulation of
electrons. Whether the negative vortex charge appears depends
solely on whether there is a sufficient AF order inside the vortex
core, as clearly seen in Fig. 3. Although our calculation is based
upon the phenomenological Hamiltonian, our results are robust
despite of different band parameters and should give a qualitative
description on the vortex physics in HTS.

We now turn to discuss the experimental results for strongly
underdoped YBa$_2$Cu$_4$O$_8$~\cite{Matsuda01} where a positively
charged vortex is reported. This seems to be inconsistent with our
prediction. We believe that the vortex charge in the above
experiment was deduced from an oversimplified assumption that the
electron density is uniform either in the absence of magnetic
field or far away from the vortex core in this strongly underdoped
HTS. In fact, experiments showed clearly the remarkable
inhomogeneities in the underdoped sample~\cite{Pan01,Underdoped}.
Many theoretical studies including the present one also show the
presence of stripe-like charge density structures in the strongly
underdoped sample~\cite{Emery,Mart00,Machida,chen02}. Upon the
application of a magnetic field, the spatial charge distribution
could become more inhomogeneous even away from the vortex core.
Therefore, their estimation of the vortex charge for the
underdoped YBa$_2$Cu$_4$O$_8$ might be invalid. For the slightly
overdoped HTS, the sample is less inhomogeneous, and their
estimation may be qualitatively correct.

With respect to the complexity of the underdoped case, we suggest
to use the spatially resolved high magnetic field NMR~\cite{NMR}
to probe the vortex charge. In this way, a  clear resolution of
the vortex core region can be reached since the fraction of the
spectrum inside the core grows with the increase of the magnetic
field.  It seems better to probe the vortex charge in slightly
underdoped samples to test our results because the strongly
underdoped samples have the complications mentioned above. We
would also like to pinpoint that the high resolution STM may be a
good candidate to probe the vortex charge by integrating local
density of states up to the chemical potential. The spatial
electron density distribution can also be directly imaged by the
electrostatic force microscope, which detects the force gradient
acting on the tip, and the scanning surface potential microscopy,
which measures the first harmonic of the force. If the vortex
indeed possesses the charge as we find here, these direct imaging
techniques can be utilized as powerful tools to study the vortex
dynamics in HTS.

We are grateful to Prof. S. H. Pan for useful discussions. This
work was supported by the Robert A. Welch Foundation, by the Texas
Center for Superconductivity at the University of Houston through
the State of Texas, and by a Texas ARP Grant (No.:
003652-0241-1999). ZDW also thanks the support from the RGC grant
of Hong Kong(HKU7092/01P) and the 973-program of Ministry of
Science and Technology of China under grant No. 1999064602.

\end{document}